\documentclass[prb,onecolumn,showpacs]{revtex4}
\usepackage{graphicx}

\begin{document}
\title{Effect of antiferromagnetic exchange interactions on the Glauber dynamics
of one-dimensional Ising models}
\author{M. G. Pini$^{1}$}
\email{mariagloria.pini@isc.cnr.it}
\author{A. Rettori$^{2,3}$}
\thanks{The authors acknowledge financial support from the
Italian Ministry for University and Research.}
\affiliation{ $^1$ Istituto dei Sistemi Complessi, CNR, Sezione di
Firenze, Via Madonna del Piano, I-50019 Sesto Fiorentino (FI),
Italy
\\
$^2$ Dipartimento di Fisica, Universit\`a di Firenze, Via G.
Sansone 1, I-50019 Sesto Fiorentino (FI), Italy
\\
$^3$ INFM-CNR National Research Center S3, Via Campi 213/A, I-41100 Modena, Italy}
\date{\today}
\begin{abstract}
We study the effect of antiferromagnetic interactions on the
single spin-flip Glauber dynamics of two different one-dimensional
(1D) Ising models with spin $\pm 1$. The first model is an Ising
chain with antiferromagnetic exchange interaction limited to
nearest neighbors and subject to an oscillating magnetic field.
The system of master equations describing the time evolution of
sublattice magnetizations can easily be solved within a linear
field approximation and a long time limit. Resonant behavior of
the magnetization as a function of temperature (stochastic
resonance) is found, at low frequency, only when spins on opposite
sublattices are uncompensated owing to different gyromagnetic
factors ({\it i.e.}, in the presence of a ferrimagnetic short
range order). The second model is the axial next-nearest neighbor
Ising (ANNNI) chain, where an antiferromagnetic exchange between
next-nearest neighbors (nnn) is assumed to compete with a
nearest-neighbor (nn) exchange
interaction of either sign. The long time response of the model to
a weak, oscillating magnetic field is investigated in the
framework of a decoupling approximation for three-spin correlation
functions, which is required to close the system of master
equations. The calculation, within such an approximate theoretical
scheme, of the dynamic critical exponent $z$, defined as ${1/
\tau} \approx ({1/ {\xi}})^z$ (where $\tau$ is the longest
relaxation time and $\xi$ is the correlation length of the chain),
suggests that the $T=0$ single spin-flip Glauber dynamics of the
ANNNI chain is in a different universality class than that of the
unfrustrated Ising chain.
\end{abstract}



\maketitle

\section{Introduction}

After the publication of fundamental
papers\cite{Parisi,Gammaitoni} on stochastic resonance (SR), it
was realized that the response amplitude of a nonlinear dynamic
system to an external periodic signal is greatly enhanced as a
function of noise strength, in the presence of a matching between
the frequency of the external force and the escape rate across an
intrinsic energy barrier. Most of the SR research\cite{reviewSR}
was pursued on dynamic systems with a double well potential,
subject to both periodic and random forces, while only a few
investigations of SR in extended or coupled systems have yet been
conducted.\cite{Sievert}

The Ising model with Glauber dynamics\cite{Glauber} can be viewed
as a set of coupled two-state oscillators, where the coherent
signal is provided by an external oscillating magnetic field and
thermal fluctuations are the only source of random noise. Each
spin is assumed to be in interaction with a heat reservoir of some
sort, which causes it to flip between the values $\sigma=+1$ and
$\sigma=-1$ randomly with time. In the presence of magnetic
coupling between the spins, the transition probability for one
spin to flip is assumed to depend on the configuration of the
neighboring spins. The time evolution of the system is described
by a master equation where the transition rates verify the
detailed-balance condition. Solving the master equation, the time
dependence of the magnetization and of the spin correlation
functions can be obtained. For exchange interaction limited to
nearest neighbor (nn) spins, the response of the Ising model with
Glauber dynamics to an oscillating magnetic field was investigated
in one (1D),\cite{Glauber,Brey_Prados} two (2D),\cite{Neda2D} and
three (3D)\cite{Neda3D,Neda23D} spatial dimensions. For the 1D nn
Ising ferromagnet, Brey and Prados\cite{Brey_Prados} obtained an
analytic expression, within the linear field approximation, for
the amplitude and the phase of the induced magnetization. The
amplitude always presents a maximum as a function of temperature,
with a genuine resonant behavior only for low frequencies.
The Glauber dynamics of the 1D Ising model with antiferromagnetic
next-nearest neighbor (nnn) exchange interaction competing with
the nn one was investigated by Yang,\cite{Yang} who employed a
decoupling approximation to solve the master equation and get an
analytical expression for the time-dependent magnetization. He
also found, by heuristic arguments, the dynamic critical exponent
$z$, defined as ${1/\tau} \approx ({1/\xi})^z$ (where $\tau$ is
the longest relaxation time and $\xi$ is the correlation length of
the chain)\cite{daSilva}, to be $z=2$, the same as that of the
unfrustrated 1D nn Ising model.

In this paper, we study - at finite temperature $T>0$ - the effect
of antiferromagnetic (AF) exchange interactions on the single
spin-flip Glauber dynamics of two different one-dimensional Ising
models. Our interest in kinetic 1D Ising models with AF
interactions is motivated by recently sinthesized
cobalt-based\cite{Caneschi,Vindigni} and
rare-earth-based\cite{Bogani,Bernot} single chain magnets, showing
slow relaxation of the magnetization at low temperature. The
magnetic properties of the former chain compound,
[Co(hfac)$_2$NITPhOMe], can be described in terms of a 1D Ising
model with AF nn exchange coupling.\cite{Vindigni,Regnault}
However, the resulting short range order is ferrimagnetic, owing
to the alternation along the chain of two different kinds of
magnetic centers (a metal ion, Co$^{2+}$, and a nitronyl-nitroxide
radical, PhOMe), both with $S=1/2$ but with different gyromagnetic
factors. In spite of further complications due to non-collinearity
of the spins,\cite{Regnault} this system was shown to be the first
experimental realization of a 1D nn Ising model with Glauber
dynamics.\cite{Vindigni} The single chain magnets belonging to the
latter class of rare-earth-based compounds, of general formula
[M(hfac)$_3$(NiTPhOPh)], where M=Eu, Gd, Tb, Dy, Ho, Er, or Yb,
and PhOPh is a nitronyl-nitroxide radical, are characterized by
strong Ising-type anisotropy and by the simultaneous presence of
both nn and nnn exchange interactions between the magnetic
centers, with the last ones being antiferromagnetic in
nature.\cite{Bogani,Bernot}

The paper is organized as follows. In Section II we investigate
the Glauber dynamics in a collinear Ising chain, with
antiferromagnetic exchange interaction limited to nearest
neighbors and different gyromagnetic factors on the two opposite
sublattices, subject to an oscillating magnetic field. The system
of master equations describing the time evolution of sublattice
magnetizations can easily be solved within a linear field
approximation and a long time limit. Resonant behavior of the
magnetization as a function of temperature (stochastic resonance)
is found, at low frequency, only when spins on opposite
sublattices are uncompensated owing to different gyromagnetic
factors ({\it i.e.}, in the presence of a ferrimagnetic short
range order). In Section III we investigate the 1D axial
next-nearest neighbor Ising (ANNNI) model, where an
antiferromagnetic exchange between next-nearest neighbor spins is
assumed to compete with a nearest-neighbor exchange
interaction of either sign. The long time response of the model to
a weak, oscillating magnetic field is investigated in the
framework of a decoupling approximation (required in order to
close the system of master equations) for three-spin correlation
functions, which in principle is  more accurate than the one
reported in Ref.~\onlinecite{Yang}. As a consequence, our
approximate calculation of the dynamic critical exponent $z$
suggests that the $T=0$ single spin-flip Glauber dynamics of the
ANNNI chain is in a different universality class than that of the
unfrustrated Ising chain. Finally, the conclusions are drawn in
Section IV.

\section{Glauber dynamics in the nearest-neighbor ferrimagnetic Ising chain}

We consider a one-dimensional Ising model with a nearest-neighbor
antiferromagnetic exchange interaction, $J<0$, in the presence of
a time-dependent external field. The Hamiltonian of the system is
\begin{equation}
\label{ferrimagnet} {\cal H}=-J\sum_{j=1}^{N}  \sigma^z_{j}
\sigma^z_{j+1} -\mu_0~ H(t)\sum_{j=1}^{N/2} (g_{A}~
\sigma^z_{2j-1}+g_{B}~ \sigma^z_{2j})
\end{equation}
where $\mu_0$ is the Bohr magneton, and $H(t)=H_0 e^{-i\omega t}$
is an external magnetic field applied along the $z$ direction and
oscillating in time with frequency $\omega$. Spins on opposite
sublattices are allowed to take possibly different gyromagnetic
factors ($g_A \ne g_B$), while we assume $\sigma^z_j=\pm 1$
$\forall j$. Hereafter, the $z$ index will be dropped for ease of
notation. In the absence of a magnetic field, if $g_A \ne g_B$ the
ground state is ferrimagnetic, with opposite uncompensated
magnetizations on the two sublattices; if $g_A=g_B$ the ground
state is antiferromagnetic, with compensated sublattice
magnetizations.

When the system is endowed with single spin-flip Glauber
dynamics,\cite{Glauber} its time evolution is described by the
master equation
\begin{equation}
\label{master} {{\partial}\over {\partial t}}p(\sigma,t)=\sum_j
\Big[ W_j(R_j \sigma) p(R_j \sigma,t) -W_j(\sigma)p(\sigma,t)
\Big]
\end{equation}
where $p(\sigma,t)$ is the probability for the system to assume
the configuration $\sigma=\{ \sigma_1, \cdots,\sigma_j,\cdots ,
\sigma_N \}$ at time $t$, $R_j \sigma$ is the configuration
obtained from $\sigma$ by flipping spin $j$, and $W_j (\sigma)$,
$W_j(R_j \sigma)$ are the transition rates between such
configurations.

For a 1D Ising model of spins ($\sigma_j=\pm 1$) with {\it ferromagnetic} nn exchange
interaction $J>0$ and gyromagnetic factor $g$, Brey and Prados\cite{Brey_Prados} showed
that, for low frequency, a stochastic resonance phenomenon occurs:
{\it i.e.}, the induced magnetization $M(t)= g \mu_0 \sum_{j=1}^N
\langle \sigma_j; t \rangle$ oscillates at the same frequency as
the magnetic field, and the amplitude of $M(t)$ presents a sharp
maximum as a function of temperature $T$. The resonance
temperature, $T_{\rm r}$, is determined by the matching between
the frequency, $\omega$, of the external field and the inverse of
the statistical time scale, $1/\tau(T_{\rm r})$, associated to the
spontaneous ({\it i.e.}, in zero field) decay of the
magnetization. In zero field, the magnetization of the 1D nn Ising
ferromagnet was found\cite{Glauber,Brey_Prados} to relax to its
equilibrium value, $M=0$, with the asymptotic $t \to \infty$
behavior $M(t)\approx e^{-t/\tau(T)}/\sqrt{t}$. The relaxation
time $\tau$ was found to be exponentially divergent for $T \to 0$,
$\tau(T) \approx e^{{4J}\over {k T}}$ (where $k$ denotes
Boltzmann's constant), and to become of the order of the inverse
of the transition rate of an isolated spin for $T\to \infty$,
$\tau \approx 1/\alpha$.\cite{Glauber,Brey_Prados}

For a 1D Ising model with {\it antiferromagnetic} nn exchange
interaction ($J<0$), the master equation (\ref{master}) is still
the starting point for the study of the chain dynamics. In this
case, if $g_A \ne g_B$, the transition rates in the presence of a
field are assumed to be different for even ($A$) and odd ($B$)
lattice sites $j$
\begin{equation}
W_j (\sigma)=W_j^{(0)}(\sigma)\Big[1 -\sigma_j \tanh (\beta_{A,B})
\Big] ={1\over 2} \alpha~ \Big[1-{1\over 2} \gamma \sigma_j
(\sigma_{j-1}+\sigma_{j+1}) \Big]~\Big[1 -\sigma_j \tanh
(\beta_{A,B}) \Big]\end{equation} \noindent where $W_j^{(0)}
(\sigma)$ denote the transition rates in zero field. The
transition rate of an isolated spin, ${1\over 2} \alpha$, is
considered as temperature independent and sets the time scale. In
the case of interacting spins, the probability per unit time of
the $j$-th spin to flip depends on the orientation of its nearest
neighbors. The magnetic field favors one orientation with respect
to the other. A correspondence between the parameters $\gamma$,
$\beta_{A,B}$ of the stochastic model and the parameters $J$,
$g_{A,B} \mu_0 H(t)$ of the statistical Ising model can be
obtained\cite{Glauber,Brey_Prados}  observing that at equilibrium
${{\partial}\over {\partial t}}p(\sigma,t)=0$, so that
\begin{equation}
\sum_j \Big[ W_j(R_j \sigma) p_{eq}(R_j \sigma,t)\Big]=\sum_j
\Big[ W_j(\sigma)p_{eq}(\sigma,t) \Big].
\end{equation}
\noindent  Next, requiring the detailed balance ({\it i.e.}, the
microscopic reversibility) condition to be satisfied
\begin{equation}
\label{detail} {{W_j (R_j \sigma)}\over {W_j(\sigma) }} =
{{p_{eq}(\sigma,t)}\over {p_{eq}(R_j \sigma,t)}},
\end{equation}
with $p_{eq}(\sigma,t)=e^{- {{{\cal H}(\sigma)}\over {k T} }}$ and
$p_{eq}(R_j \sigma,t)=e^{-{{{\cal H}(R_j \sigma)}\over {k T}}},$
one readily obtains
\begin{equation}
\gamma=\tanh \Big({{2J}\over {k T}} \Big),~~~ \beta_{A,B}=\tanh
\Big( { {g_{A,B} \mu_0 H(t)}\over {k T}} \Big).
\end{equation}

The evolution equation for the spin expectation value
$\langle\sigma_j;t \rangle=\sum_{\sigma} \sigma_j p(\sigma,t)$ is
directly obtained from the master equation to be ${{\partial}\over
{\partial t}}\langle \sigma_{j}; t\rangle=-2\langle \sigma_j
W_j(\sigma); t \rangle$.\cite{Glauber,Brey_Prados} Considering
that for model (\ref{ferrimagnet}) the spins belong to two
opposite sublattices, the system of evolution equations in the
presence of an oscillating field is
\begin{eqnarray}
{{\partial}\over {\partial (\alpha t)}}\langle \sigma_{2j-1};
t\rangle&=&  -\langle \sigma_{2j-1}; t\rangle + {1\over 2} \gamma
\Big( \langle \sigma_{2j-2}; t\rangle +\langle \sigma_{2j};
t\rangle \Big) + \beta_A \Big[ 1 -{1\over 2} \gamma \Big( \langle
\sigma_{2j-2}\sigma_{2j-1}; t\rangle + \langle
\sigma_{2j-1}\sigma_{2j}; t\rangle \Big)\Big] \cr {{\partial}\over
{\partial (\alpha t)}}\langle \sigma_{2j}; t \rangle&=& -\langle
\sigma_{2j}; t\rangle + {1\over 2} \gamma \Big( \langle
\sigma_{2j-1}; t\rangle +\langle \sigma_{2j+1}; t\rangle \Big) +
\beta_B \Big[ 1 -{1\over 2} \gamma \Big( \langle
\sigma_{2j-1}\sigma_{2j}; t\rangle + \langle
\sigma_{2j}\sigma_{2j+1}; t\rangle \Big)\Big]
\end{eqnarray}

The system is not closed owing to the presence of two-spin,
time-dependent correlation functions on the right hand sides. In
order to solve it, a linear field approximation is
made\cite{Glauber,Brey_Prados} so that $\tanh [(g_{A,B} \mu_0
H_0)/(k T)]$ can be expanded for small values of the argument and
two-spin correlations can be evaluated in the absence of a field.
Moreover, if in the long time limit $t \to \infty$ the nn
correlation functions are assumed\cite{Glauber,Brey_Prados} to
take their equilibrium value $\langle \sigma_j \sigma_{j+1};
t\rangle \to \eta=\tanh[J/(k T)]$, the system of two coupled
equations of motion for the two sublattice magnetizations
\begin{equation} \label{dueperdue} M_1(t)=g_A \mu_0 \sum_{j=1}^{N/2} \langle \sigma_{2j-1};
t\rangle,~~~ M_2(t)=g_B \mu_0 \sum_{j=1}^{N/2} \langle
\sigma_{2j}; t\rangle \end{equation} can be written in matrix form

\[ \left( \begin{array}{c}{{\partial }\over {\partial (\alpha t)}}M_1(t) \\ {{\partial }\over {\partial (\alpha t)}}M_2(t) \end{array} \right)
+\left( \begin{array}{cc} 1 & -{{g_A}\over {g_B}}
\gamma \\
-{{g_B}\over {g_A}} \gamma & 1  \end{array} \right)\left(
\begin{array}{c}
M_1(t) \\ M_2(t) \end{array} \right)={\cal N}(T) \left(
\begin{array}{c}
g_A^2 \\ g_B^2 \end{array} \right)e^{-i \omega t}
\]
\begin{equation}
\label{twotwo}
\end{equation}
\noindent Taking into account that $\gamma={{2\eta}\over
{1+\eta^2}}$, the temperature dependent coefficient ${\cal N}(T)$
can be expressed as
\begin{equation}
{\cal N}(T)={N\over 2} {{\mu_0^2 H_0}\over {k T}} (1-\gamma
\eta)={N\over 2} {{\mu_0^2 H_0}\over {k T}} {{1-\eta^2}\over
{1+\eta^2}}.
\end{equation}
The above system can be decoupled diagonalizing the $2 \times 2$
non-symmetric matrix on the l.h.s. of Eq.~(\ref{twotwo}). Denoting
by ${\cal M}_1(t)$ and ${\cal M}_2(t)$ the normal modes, one
obtains
\[ \left( \begin{array}{c}{{\partial }\over
{\partial (\alpha t)}}{\cal M}_1(t) \\
{{\partial }\over {\partial (\alpha t)}}{\cal M}_2(t) \end{array}
\right) +\left( \begin{array}{cc} \lambda_1
 & 0 \\
0 & \lambda_2  \end{array} \right)\left(
\begin{array}{c}
{\cal M}_1(t) \\ {\cal M}_2(t) \end{array} \right)={\cal N}(T)
 \left(
\begin{array}{c}f_1
\\ f_2 \end{array} \right)e^{- i \omega t}
\]
\noindent where the eigenvalues $\lambda_n$ ($n=1,2$) turn out to
be independent of the gyromagnetic factors $g_A$ and $g_B$
\begin{equation} \lambda_1=1-\gamma, ~~~~\lambda_2=1+\gamma,
\end{equation}
and the $f_n$ ($n=1,2$) coefficients are
\begin{equation} f_1={{g_B}\over 2}(g_B+g_A)~~~~f_2={{g_B}\over 2}
(g_B-g_A).\end{equation}

The relationships between the normal modes ${\cal M}_n(t)$ and the
sublattice magnetizations $M_n(t)$ ($n=1,2$) are
\begin{equation}
\label{calMM2} {\cal M}_1(t)={1\over 2}\Big[M_2(t)+{{g_B}\over
{g_A}} M_1(t)\Big],~~~{\cal M}_2(t)={1\over
2}\Big[M_2(t)-{{g_B}\over {g_A}} M_1(t)\Big]
\end{equation}
({\it i.e.}, ${\cal M}_1(t)$ and ${\cal M}_2(t)$ are related to the net and
the staggered magnetization, respectively). Conversely, one has
\begin{equation}
\label{McalM2} M_1(t)={{g_A}\over {g_B}}\Big[ {\cal M}_1(t)-{\cal
M}_2(t) \Big],~~~ M_2(t)={\cal M}_1(t)+{\cal M}_2(t).
\end{equation}

\noindent The general solution for the normal modes is ($n=1,2$)
\begin{equation}
\label{general} {\cal M}_n(t)={\cal M}_n(t_0)e^{-{{t-t_0}\over
{\tau_n}}}+{\cal N}(T) f_n\int_{t_0}^t dt^{\prime}
e^{{{t^{\prime}-t}\over {\tau_n}}} e^{-i \omega t^{\prime}}
\end{equation}
\noindent where the relaxation times $\tau_n$ are expressed, in
terms of the eigenvalues $\lambda_n$ of the non-symmetric $2
\times 2$ matrix, as $\tau_n=1/(\alpha \lambda_n)$, so that
\begin{equation}\label{tempi} \tau_1={1\over
{\alpha(1-\gamma)}},~~~\tau_2={1\over{\alpha(1+\gamma)}}.
\end{equation}

In the absence of an external magnetic field, ${\cal N}(T)=0$,
the normal modes ${\cal M}_n(t)$ are found to relax exponentially.
In the low temperature limit, $T \to 0$, one has
$\gamma=\tanh\Big( {{2J}\over {k T}}\Big)\approx {{J}\over {\vert
J \vert }} \Big(1-2e^{-{{4\vert J\vert}\over {kT}}} \Big)$, so
that for antiferromagnetic nn exchange ($J<0$), the first
relaxation time is simply $\tau_1 \approx {1\over {2\alpha}}$,
while the second relaxation time is exponentially diverging with
decreasing $T$, $\tau_2 \approx {1\over {2\alpha}}e^{{{4\vert
J\vert}\over {k T}}}$. For high temperatures, $k T \gg \vert J
\vert$, both relaxation times become of the order of the inverse
of the transition rate of an isolated spin, $\tau_1 \approx \tau_2
\approx 1/\alpha$.

For non vanishing magnetic field, the time dependence of the
normal modes is obtained letting $t_0 \to -\infty$ in
Eq.~(\ref{general})
\begin{equation}
{\cal M}_n(t)={\cal N}(T)  {{f_n}\over {\lambda_n}}{{1}\over {1-i
\omega \tau_n}} e^{- i \omega t}~~~~(n=1,2)
\end{equation}

The total magnetization is
\begin{equation}
M_{tot}(t)=M_1(t)+M_2(t)={{g_B+g_A}\over {g_B}} {\cal
M}_1(t)+{{g_B-g_A}\over {g_B}} {\cal M}_2(t)=\chi(\omega,T) H_0
e^{-i \omega t},\end{equation} \noindent where the complex
susceptibility $\chi(\omega,T)$ is given by
\begin{equation}
\label{susc} \chi(\omega,T)=N ~{{\mu_0^2}\over {k T}}~ \Bigg[
\Big({{g_B + g_A}\over 2}\Big)^2~{{1+\eta}\over
{1-\eta}}~{{1}\over {1-i \omega \tau_1}}+\Big({{g_B -g_A}\over 2}
\Big)^2~{{1-\eta}\over {1+\eta}}~{{1}\over {1-i \omega
\tau_2}}\Bigg].
\end{equation}

\begin{figure}
\includegraphics[width=16cm,angle=0,bbllx=109pt,bblly=310pt,%
bburx=550pt,bbury=597pt,clip=true]{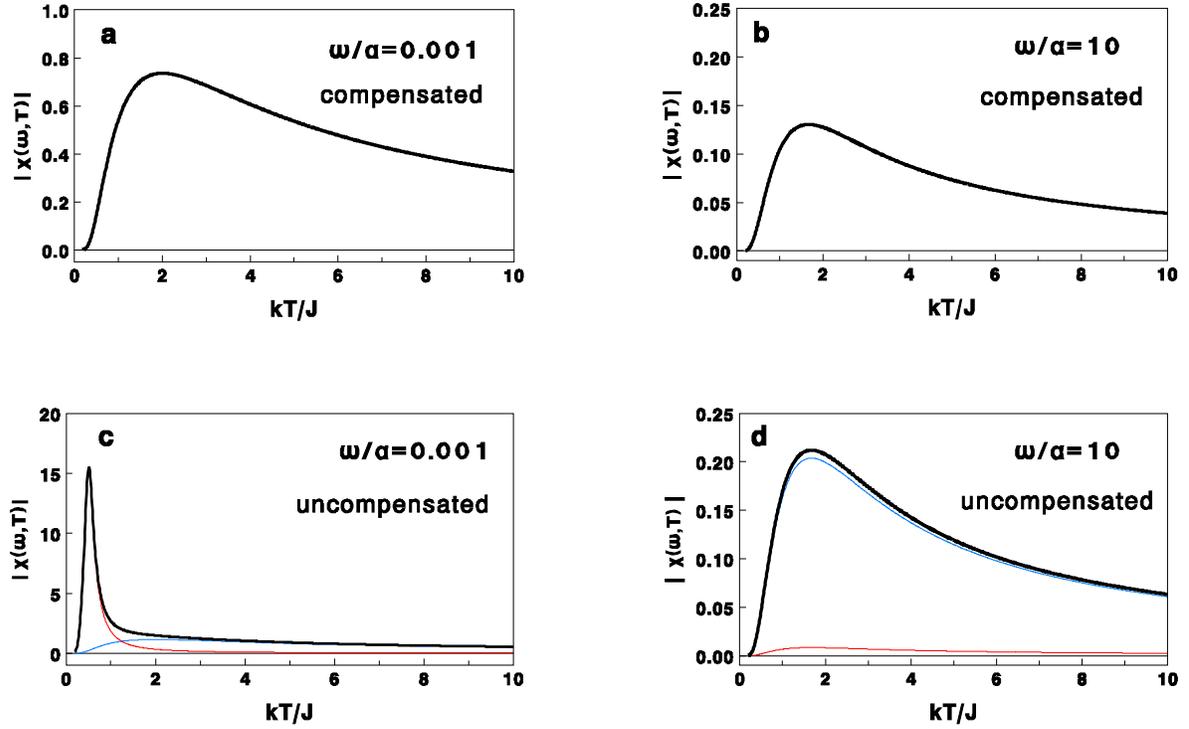} \caption{(color
online) Temperature dependence of the amplitude of the complex
susceptibility $\vert \chi(\omega,T) \vert$ for an Ising chain
with antiferromagnetic nearest neighbor interaction $J=-1$,
subject to a weak external magnetic field oscillating at frequency
$\omega$. Figures (a), (b) refer to the compensated case
($g_A=g_B=2$), while Figures (c) and (d) to the uncompensated case
($g_A=2,~g_B=3$), for selected values of the frequency ($\omega/
\alpha=0.001$ and $10$). In Figs.~(c,d) the thin (color) lines
represent the contributions to the amplitude of the two terms on
the r.h.s. of Eq.~(\ref{susc}), while the thick (black) line is
their sum. A resonant behavior (similar to the one predicted for
the nn Ising ferromagnetic chain endowed with single spin-flip
Glauber dynamics)\cite{Glauber,Brey_Prados} is observed only in
the uncompensated case for low frequency (notice the enhanced
vertical scale in Fig.~c). }
\end{figure}

\begin{figure}
\includegraphics[width=12cm,angle=0,bbllx=53pt,bblly=234pt,%
bburx=557pt,bbury=541pt,clip=true]{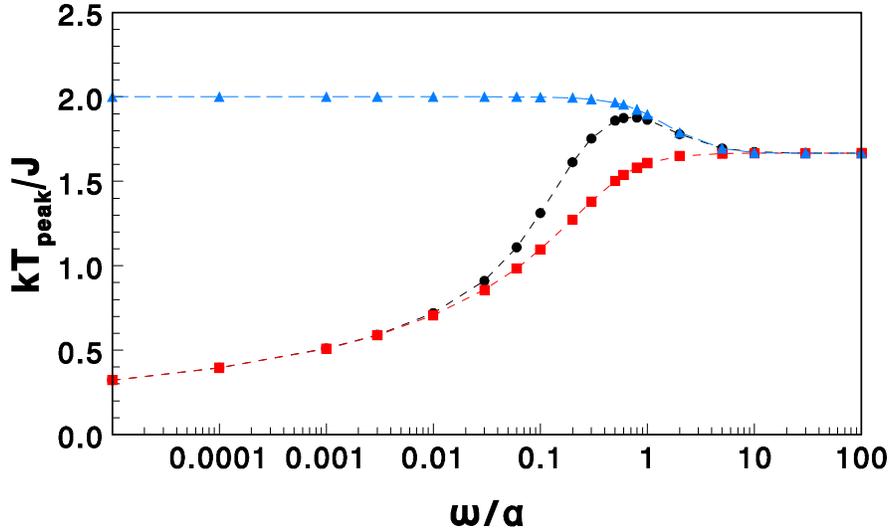} \caption{(color
online) Frequency dependence of the peak temperature of the
amplitude of the complex susceptibility $\vert \chi(\omega,T)
\vert$ of an Ising chain with nearest neighbor exchange
interaction. Triangles: compensated antiferromagnet ($J=-1$,
$g_A=g_B=2$); circles: uncompensated ferrimagnet ($J=-1$, $g_A=2$,
$g_B=3$); squares: ferromagnet ($J=+1$,
$g_A=g_B=2$).\cite{Brey_Prados} The dashed lines are guides to the
eye.}
\end{figure}

In the limit $\omega \to 0$, the static susceptibility of the
Ising ferrimagnetic chain in zero field is correctly recovered:
see Appendix A.1, Eq.~(\ref{chisAF}), for details. As regards the
dynamic response of the system to a weak, oscillating magnetic
field, from Eq.~(\ref{susc}) it is apparent that, for
antiferromagnetic nn exchange ($J<0$) and $T \to 0$, the first
term on the r.h.s. is associated with a fast relaxation, while the
second term with an exponentially slow relaxation. Thus, a
resonant behavior, similar to the one observed in the
ferromagnetic nn Ising chain endowed with single spin-flip Glauber
dynamics,\cite{Brey_Prados} is possible only when spins on
opposite sublattices are uncompensated owing to different
gyromagnetic factors ({\it i.e.}, in the presence of ferrimagnetic
short range order). See Fig.~1, where the temperature dependence
of the amplitude of $\chi(\omega,T)$ is reported, for selected
values of the frequency, both in the compensated ($J<0$ and
$g_A=g_B$) and uncompensated ($J<0$ and $g_A \ne g_B$) case.

The resonant behavior shown by the ferrimagnetic chain at low
frequency (see Fig.~1c) is a manifestation of the stochastic
resonance phenomenon:\cite{reviewSR} {\it i.e.}, the response of a
set of coupled bistable systems to a periodic drive is enhanced in
the presence of a stochastic noise when a matching occurs between
the fluctuation induced switching rate of the system and the
forcing frequency. In the ferrimagnetic chain, the role of
stochastic noise is played by thermal fluctuations and the
resonance peak occurs when the deterministic time scale of the
external magnetic field matches with the statistical time scale
associated to the spontaneous decay of the net magnetization
$M_{tot}(t)$. For low frequency $\omega \ll \alpha$ ({\it i.e.},
low temperature), the resonance condition for the uncompensated
case is
\begin{equation}
\label{resonance_condition} \omega^{-1} \approx \tau_2(T_{peak}),
\end{equation}
while for the compensated case only the mode with fast relaxation
$\tau_1 \approx O(\alpha^{-1})$ contributes, providing  a broad
peak rather than a genuine resonance. For high frequency $\omega
\gg \alpha$ ({\it i.e.}, high temperature) a broad peak is found,
both for the uncompensated and the compensated case, since the two
relaxation times $\tau_1$ and $\tau_2$ become of the order of
$1/\alpha$, so that the resonance condition cannot be
fulfilled.\cite{Brey_Prados}

The frequency dependence of the peak temperature $T_{peak}$ is
reported in Fig.~2 both for the compensated (antiferromagnetic)
and the uncompensated (ferrimagnetic) chain, and compared with the
ferromagnetic counterpart.\cite{Brey_Prados} In the compensated
case, the frequency dependence of the peak is very smooth, owing
to the smooth temperature dependence of the relaxation time
$\tau_1$, ranging between $1/(2\alpha)$ at low $T$ and $1/\alpha$
at high $T$. In the uncompensated case, a behavior very similar to
the ferromagnetic one is observed for low frequency: the reason is
that for low $\omega$ the dominant contribution to $\chi(\omega)$
is provided by the second term on the r.h.s. of Eq.~(\ref{susc}).
At intermediate frequency, a maximum is observed owing to the
coming into play of the first term on the r.h.s. of
Eq.~(\ref{susc}). Finally, for $\omega \gg \alpha$, the amplitude
of $\chi(\omega)$ becomes
\begin{equation}
\label{suschigh} \vert \chi(\omega,T) \vert  \approx N~
{{\mu_0^2}\over {k T}} ~{{\alpha}\over {\omega}}~\Bigg[ \Big({{g_B
+ g_A}\over 2}\Big)^2~{{1+\eta}\over
{1-\eta}}~(1-\gamma)+\Big({{g_B -g_A}\over
2}\Big)^2~{{1-\eta}\over {1+\eta}}~(1+\gamma)\Bigg]
\end{equation}
\noindent where both terms in square brackets on the r.h.s. of
Eq.~(\ref{suschigh}) present a maximum at the same temperature,
which is numerically determined to be $T_{peak} \approx 1.66711
\vert J \vert$.

\section{Glauber dynamics in the axial-next-nearest-neighbor-Ising (ANNNI) chain}

We consider a 1D axial-next-nearest-neighbor Ising (ANNNI) model
with spins alternating on two interlacing sublattices (denoted by
$A$ and $B$), with Hamiltonian
\begin{equation}
\label{ANNNI} {\cal H}=-J_1\sum_{i=1}^{N/2}  \Big(\sigma^z_{2i-1}
\sigma^z_{2i} + \sigma^z_{2i} \sigma^z_{2i+1}\Big) -J_{2}
\sum_{i=1}^{N/2} \Big(\sigma^z_{2i-1} \sigma^z_{2i+1}+
\sigma^z_{2i} \sigma^z_{2i+2}\Big)-\mu_0 H(t)\sum_{i=1}^{N/2}
\Big(g_{A} \sigma^z_{2i-1}+g_{B} \sigma^z_{2i}\Big).
\end{equation}
The intra-sublattice antiferromagnetic next-nearest neighbor
coupling $J_{2}<0$ competes with the inter-sublattice nearest
neighbor coupling $J_1$, which may be of either sign. In what
follows, we shall assume $J_1>0$ (ferromagnetic coupling).
$H(t)=H_0 e^{i\omega t}$ is an external magnetic field applied
along the $z$ direction and oscillating in time with frequency
$\omega$, $\mu_0$ denotes the Bohr magneton and the spins
$\sigma^z_i=\pm 1$ are allowed to assume possibly different
gyromagnetic factors on odd and even sites ($g_A \ne g_B$); the
$z$ index shall be dropped for ease of notation.

In the limiting case $g_A=g_B=g$, Eq. (1) reduces to the
well-known ANNNI (axial next-nearest-neighbor Ising)
model.\cite{Selke} Depending on the competition ratio
$r=-J_2/J_1$, this model in zero field is known to admit a
ferromagnetic ground state for $r<1/2$, and a $(2,2)$ antiphase
structure (two spins up, two spins down), with zero magnetization,
for $r>1/2$; for $r=1/2$ the ground state is degenerate and
disordered.\cite{Morita_gs} At finite temperatures, the model
cannot support long range order; however, a strong short range
order is present in the paramagnetic phase. For zero applied
field, as far as the thermodynamic properties are
concerned,\cite{notaYang} the 1D ANNNI model can be mapped into an
equivalent 1D Ising model with only nearest neigbor interaction in
an effective field, and analytic results (see Appendix A.2) can be
obtained for the partition function and the spin correlation
functions.\cite{Stephenson,Harada} In the presence of a static
magnetic field, the ground state of the generalized ANNNI model,
{\it i.e.} a chain of alternating spins with different quantum
numbers and different nnn exchange interactions on the two
sublattices, was thoroughly investigated,\cite{Kim_gs} and the
thermodynamic properties were exactly calculated (though
numerically) by the transfer matrix method.\cite{PR_tm,Kim_tm}

Here we aim at investigating the long-time dynamic response of the
ANNNI chain, Eq.~(\ref{ANNNI}), to a weak, external magnetic field
oscillating in time. The time evolution of the system is still
described by the master equation (\ref{master}), but with respect
to the case of the nn Ising chain, the transition rates in zero
field, $W_j^{(0)} (\sigma)$, are now assumed to take the form
\begin{equation}
\label{zeroANNNI} W_j^{(0)} (\sigma)={1\over 2} \alpha \Big[ 1-
{1\over 2} \gamma_1 \sigma_j (\sigma_{j-1}+\sigma_{j+1})\Big]
\Big[1-{1\over 2} \gamma_2 \sigma_j
(\sigma_{j-2}+\sigma_{j+2})\Big]
\end{equation}
meaning that the probability per unit time of the $j$-th spin to
flip depends on the status of both its nearest neighbors and next
nearest neighbors; ${1\over 2} \alpha$, the transition rate of an
isolated spin, is arbitrary and sets the time scale. In the
presence of a field applied along the $z$ axis, the transition
rates $W_j (\sigma)$ are given by
\begin{equation}
W_j (\sigma)=W_j^{(0)}(\sigma)\Big[1 -\sigma_j \tanh (\beta_{A,B})
\Big]. \end{equation}

As usual, a correspondence between the parameters $\gamma_1$,
$\gamma_{2}$, $\beta_{A,B}$ of the stochastic model and the
parameters $J_1$, $J_{2}$, $g_{A,B} \mu_0 H(t)$ of the statistical
ANNNI model can be obtained requiring the detailed balance ({\it
i.e.}, the microscopic reversibility) condition,
Eq.~(\ref{detail}) to be satisfied at equilibrium. One finds\cite{Yang}
\begin{equation}
\gamma_1=\tanh \Big({{2J_1}\over {k T}} \Big),~~~\gamma_2=\tanh
\Big( {{2 J_{2}} \over {k T}} \Big),~~~\beta_{A,B}=\tanh \Big( {
{g_{A,B} \mu_0 H(t)}\over {k T}} \Big).
\end{equation}

The stochastic equation of motion for the spin expectation value
$\langle\sigma_j;t \rangle=\sum_{\sigma} \sigma_j p(\sigma,t)$ in
the presence of an oscillating field is then obtained,
from the master equation, to be ${{\partial}\over {\partial t}}\langle
\sigma_{j}; t\rangle=-2\langle \sigma_j W_j(\sigma); t
\rangle$, giving\cite{Yang}
\begin{eqnarray}
\label{motion} {{\partial}\over {\partial t}}\langle\sigma_j;t
\rangle&=&-\langle \sigma_j;t\rangle + {1\over 2} \gamma_1 \Big(
\langle\sigma_{j-1};t\rangle+\langle\sigma_{j+1};t\rangle \Big) +
{1\over 2} \gamma_{2} \Big(
\langle\sigma_{j-2};t\rangle+\langle\sigma_{j+2};t\rangle \Big)\cr
&-&{1\over 4}\gamma_1 \gamma_{2} \Big( \langle
\sigma_j\sigma_{j-1}\sigma_{j-2};t \rangle+ \langle \sigma_j
\sigma_{j-1} \sigma_{j+2};t \rangle + \langle \sigma_j
\sigma_{j+1}\sigma_{j-2};t \rangle+ \langle \sigma_j \sigma_{j+1}
\sigma_{j+2};t \rangle \Big)\cr &+&\tanh \Big( {{g_{A,B} \mu_0
H(t)}\over {k T}}\Big)\cr &\times&\Big[ 1-{1\over 2} \gamma_1
\Big(\langle \sigma_j\sigma_{j-1};t \rangle +\langle \sigma_j
\sigma_{j+1};t\rangle \Big) -{1\over 2} \gamma_{2} \Big(\langle
\sigma_j \sigma_{j-2};t\rangle+\langle \sigma_j
\sigma_{j+2};t\rangle \Big) \cr &+&{1\over 4} \gamma_1 \gamma_{2}
\Big(\langle \sigma_{j-1}\sigma_{j-2};t\rangle +\langle
\sigma_{j-1}\sigma_{j+2};t\rangle +\langle
\sigma_{j+1}\sigma_{j-2};t\rangle +\langle
\sigma_{j+1}\sigma_{j+2};t\rangle \Big) \Big].
\end{eqnarray}
\noindent where we remind that the subscripts $A$ and $B$ refer to
the case of $j$ odd and $j$ even, respectively. This set of
equations is not closed, owing to the time-dependent two-spin and
three-spin correlation functions on the r.h.s. In order to solve
it, we make the following approximations.
\begin{itemize}
\item For sufficiently weak fields ($x=(g_{A,B} \mu_0 H_0)/(k T)\ll 1$),
the hyperbolic tangent on the r.h.s. of Eq.~(\ref{motion}) is
expanded for low values of the argument ($\tanh x  \approx x$) and
two-spin correlation functions are calculated in the absence of a
field.
\item Three-spin correlation functions are decoupled, {\it in all possible ways},
into products of a single-spin expectation value and a two-spin
correlation function
\begin{equation} \label{decoupling} \langle \sigma_j \sigma_{j+m} \sigma_{j+n}; t\rangle
\approx \langle \sigma_j;t\rangle \langle
\sigma_{j+m}\sigma_{j+n}; t\rangle  +\langle \sigma_{j+m};t\rangle
\langle \sigma_j \sigma_{j+n} ; t\rangle  + \langle
\sigma_{j+n};t\rangle \langle \sigma_j\sigma_{j+m}; t\rangle.
\end{equation}
\noindent Notice that a different, and incomplete, decoupling was
adopted in Ref.~\onlinecite{Yang}, thus leading to different
results with respect to the present work.

\item For sufficiently long times, two-spin
correlation functions between $n$-th neighbors are assumed to be
independent of the initial conditions and to take their static
equilibrium values $\langle \sigma_j \sigma_{j+n}; t \rangle \to
\eta_n$ for $t\to \infty$. Static two-spin correlation functions
$\eta_n=\langle \sigma _j \sigma_{j+n} \rangle$ can be exactly
calculated in 1D via the transfer matrix
method.\cite{Stephenson,Harada,PR_tm,Kim_tm} For $g_A=g_B$ and
$H_0=0$, analytic results\cite{Stephenson,Harada} can be obtained
for $\eta_n$: see Appendix A.2 for details.
\end{itemize}

Under these approximations, the master equation for the spin
expectation value on a generic site $j$ becomes
\begin{eqnarray}
\label{generic} {{\partial}\over {\partial (\alpha
t)}}\langle\sigma_j;t \rangle&=&- \Big[1+ {1\over 2} \gamma_1
\gamma_{2} (\eta_1+\eta_3) \Big]\langle \sigma_j;t\rangle \cr &+&
{1\over 2} \gamma_1 \Big( 1 - \gamma_{2} \eta_{2} \Big) \Big[
\langle\sigma_{j-1};t\rangle + \langle\sigma_{j+1};t\rangle\Big]
+{1\over 2} \gamma_{2} \Big(1-\gamma_1 \eta_1 \Big) \Big[
\langle\sigma_{j-2};t\rangle +\langle\sigma_{j+2};t\rangle\Big]
\cr &+&\Big[ {{g_{A,B} \mu_0 H(t)}\over {k T}}\Big]~ \Big[ 1-
\gamma_1 \eta_1 -\gamma_{2} \eta_{2} +{1\over 2} \gamma_1
\gamma_{2} (\eta_1 +\eta_3)\Big] .
\end{eqnarray}

In the range of the competition ratio $r$ corresponding to weak
nnn antiferromagnetism $0<r<{1\over 2}$, the ground state of the
model is ferromagnetic (since we have assumed $J_1>0$), while for
strong nnn antiferromagnetism ${1\over 2}<r<\infty$, the ground
state is the so-called $(2,2)$ antiphase state, consisting of two
spins up followed by two spins down. The two different regimes
shall be investigated separately since they require different
order parameters.

\subsection{Weak nnn antiferromagnetism (competition ratio $0<r<{1\over 2}$)}

In the range of the competition ratio $r$ corresponding to the
ferromagnetic ground state ($0<r<{1\over 2}$), owing to the
different gyromagnetic factors on odd ($g_A$) and even ($g_B$)
lattice sites, it is necessary to consider the magnetizations over
{\it two} sublattices, like in Eq.~(\ref{dueperdue}), as the order
parameter. From the master equation, Eq.~(\ref{generic}), one is
thus led to consider a system of two coupled equations of motion,
which can be written just like Eq.~(\ref{twotwo}), with the
elements of the $2 \times 2$ non-symmetric matrix now being
\begin{eqnarray}
a_{11}&=&1-\gamma_2(1-\gamma_1 \eta_1)+{1\over 2}\gamma_1
\gamma_2(\eta_1+\eta_3)=a_{22}\cr a_{12}&=&-{{g_A}\over {g_B}}
\gamma_1 (1-\gamma_2 \eta_2)~~~a_{21}=-{{g_B}\over {g_A}} \gamma_1
(1-\gamma_2 \eta_2)
\end{eqnarray}
\noindent and the temperature dependent coefficient
\begin{equation}
{\cal N}(T)={N\over 2} {{\mu _0^2 H_0}\over {k T}} \Big[
1-\gamma_1 \eta_1 -\gamma_2 \eta_2 +{1\over 2} \gamma_2 \gamma_1
(\eta_1+\eta_3) \Big].
\end{equation}
After diagonalization, the eigenvalues now turn out to be
\begin{eqnarray}
\label{eigenJ1J2} \lambda_1&=& 1-\gamma_1
(1-\gamma_2\eta_2)-\gamma_2(1-\gamma_1\eta_1)+{1\over
2}\gamma_1\gamma_2(\eta_1+\eta_3)\cr \lambda_2&=&1+\gamma_1
(1-\gamma_2\eta_2)-\gamma_2(1-\gamma_1\eta_1)+{1\over
2}\gamma_1\gamma_2(\eta_1+\eta_3),
\end{eqnarray}
\noindent independent of the gyromagnetic factors $g_A$ and $g_B$.
The relationships between the normal modes ${\cal M}_n(t)$ and the
sublattice magnetizations $M_n(t)$ ($n=1,2$) are the same as in
Eqs.~(\ref{calMM2}), (\ref{McalM2}). Also the expressions for the
$f$-coefficients are the same, {\it i.e.} $f_1={{g_B}\over
2}(g_B+g_A)$, $f_2={{g_B}\over 2} (g_B-g_A)$. As before, the
general solution for the normal modes takes the form in
Eq.~(\ref{general}), where the relaxation times are
$\tau_n={1\over {\alpha \lambda_n}}$, with the eigenvalues now
given by Eq.~(\ref{eigenJ1J2}). Finally, in the case of weak nnn
antiferromagnetic coupling, the complex susceptibility of the
ANNNI chain turns out to be
\begin{eqnarray}
\label{suscANNNI2} \chi(\omega,T)&=&N ~{{\mu_0^2}\over {k
T}}~\Big[ 1-\gamma_1 \eta_1 -\gamma_2 \eta_2 +{1\over 2} \gamma_1
\gamma_2 (\eta_1+\eta_3) \Big]\cr &\times&
\Bigg[\Big({{g_B+g_A}\over 2}\Big)^2 {1\over {\lambda_1}} {1\over
{1-i\omega \tau_1}}+ \Big({{g_B-g_A}\over 2}\Big)^2 {1\over
{\lambda_2}} {1\over {1-i\omega \tau_2}}\Bigg].
\end{eqnarray}

\begin{figure}
\includegraphics[width=12cm,angle=0,bbllx=67pt,bblly=220pt,%
bburx=552pt,bbury=549pt,clip=true]{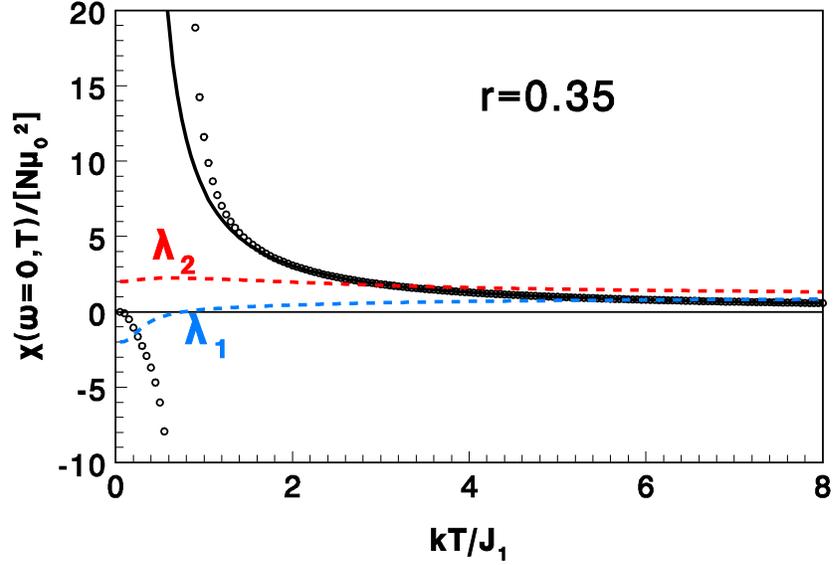} \caption{(color
online) Temperature dependence of the static susceptibility
$\chi(\omega=0,T)$ of an ANNNI chain with $J_1=1$, $J_2=-0.35$ and
$g_A=g_B=2$, corresponding to a value $r=0.35$ of the competition
ratio (weak nnn antiferromagnetism). The thick line is the exact
transfer matrix result, while open circles denote the approximate
calculation, Eq.~(\ref{suscANNNI2}). The temperature dependence of
the eigenvalues $\lambda_1$ and $\lambda_2$ (\ref{eigenJ1J2}) is
also shown by the dashed lines. The approximations made to close
the set of master equations (\ref{motion}) are found to fail for
low temperatures.}
\end{figure}

In the limiting case $r=0$, the well-known result for the nn Ising
chain\cite{Glauber,Brey_Prados} is correctly recovered. In the
case $0<r<{1\over 2}$, we show in Fig.~3 that the approximate
static susceptibility, calculated from Eq.~(\ref{suscANNNI2}) for
zero frequency, turns out to be in good agreement with the exact
transfer matrix result,\cite{Stephenson,Harada}
Eq.~(\ref{suscSteph}), only at high temperatures ($k T \gtrsim
J_1$). In contrast, an unphysical (negative) static susceptibility
is obtained at low temperatures, as a consequence of the negative
values assumed by the eigenvalue $\lambda_1$ for $k T \lesssim
J_1$.

The low-temperature failure of Eq.~(\ref{suscANNNI2}) can be
attributed to the decoupling (\ref{decoupling}) of three-spin
correlation functions, which was made in order to close the set of
master equations, Eq.~(\ref{motion}): in fact, decoupling
approximations have the drawback to be uncontrollable, but in
principle they are expected to be more accurate the higher the
temperature. Moreover, at low temperatures one can guess another
source of error to lie in the assumption that, for sufficiently
long times, the spin-spin correlation functions take their {\it
static} equilibrium values: $\langle \sigma_j \sigma_{j+n}; t
\rangle \to \eta_n$ for $t\to \infty$. In fact, for competition
ratio in the range $0<r<1$, the 1D ANNNI model with Glauber
dynamics is known to be lacking in ergodicity at $T=0$: the ground
state can {\it not} be reached by single spin-flip Glauber
dynamics, after a sudden cooling of the system down to $T=0$
starting from high temperature. The difference between the static
($r=1/2$)\cite{Morita_gs} and the dynamic ($r=1$)\cite{Redner}
ground state phase boundary of the 1D ANNNI model was pointed out
by Redner and Krapivsky\cite{Redner}, who showed that for
$0<r<1/2$ the ferromagnetic ground state can not be reached
because of the repulsion between domain walls which forces them to
be at least two lattice constants apart, while for $1/2<r<1$ the
$(2,2)$ antiphase ground state can not be reached owing to the
persistence of isolated domains of length $\ge 3$.\cite{Redner} In
contrast, both for $r=0$ (1D nn Ising model)\cite{Glauber,Bray}
and $r>1$ (1D ANNNI model with strong nnn AF
coupling)\cite{Redner} the ground state can asymptotically ($t \to
\infty$) be reached at $T=0$.

The low temperature failure of our approximate theory in the case
$0<r<{1\over 2}$ prevented us from calculating the temperature
dependence of the amplitude of the complex susceptibility.
However, it is worth observing that, since for $T \to 0$ the
zero-field static susceptibility diverges,\cite{Stephenson} a
resonant behavior might be expected for low frequency provided
that the system admits also a diverging relaxation time for low
temperature.

\subsection{Strong nnn antiferromagnetism (competition ratio ${1\over 2}<r< \infty$)}

In the range of the competition ratio $r$ corresponding to the
$(2,2)$-antiphase state (${1\over 2}<r<\infty$), it is necessary
to consider the magnetizations over {\it four}
sublattices\cite{Sen}
\begin{eqnarray}
M_1(t)&=&g_A \mu_0 \sum_{j=0}^{{N\over 4}-1} \langle
\sigma_{1+4j}; t\rangle,~~~M_2(t)=g_B \mu_0 \sum_{j=0}^{{N\over
4}-1} \langle \sigma_{2+4j}; t\rangle \cr M_3(t)&=&g_A \mu_0
\sum_{j=0}^{{N\over 4}-1} \langle \sigma_{3+4j}; t\rangle,~~~
M_4(t)=g_B \mu_0 \sum_{j=0}^{{N\over 4}-1} \langle \sigma_{4+4j};
t\rangle
\end{eqnarray}
as the order parameter. One is thus led to consider a system of
four coupled equations of motion, which can be written in matrix
form as

\[ \left( \begin{array}{c}
{{\partial }\over {\partial (\alpha t)}}M_1(t)
\\
{{\partial }\over {\partial (\alpha t)}}M_2(t)
\\
{{\partial }\over {\partial (\alpha t)}}M_3(t)
\\
{{\partial }\over {\partial (\alpha t)}}M_4(t)
\end{array} \right)
+\left( \begin{array}{cccc} {\cal A} & {\cal B} & {\cal C} & {\cal
B} \\
{\cal D} & {\cal A} & {\cal D} & {\cal C}
\\
{\cal C} & {\cal B} & {\cal A}& {\cal B}
\\ {\cal D} & {\cal C} & {\cal D}  & {\cal A}
\end{array} \right)\left(
\begin{array}{c}
M_1(t) \\ M_2(t) \\ M_3(t) \\ M_4(t) \end{array} \right)={\cal
N}(T) \left(
\begin{array}{c}
g_A^2 \\ g_B^2 \\g_A^2 \\ g_B^2\end{array} \right)e^{-i \omega t}
\]

\noindent where
\begin{eqnarray}
\nonumber {\cal A}&=& a_{11}=a_{22}=a_{33}=a_{44}=1+{1\over
2}\gamma_2 \gamma_1 (\eta_1+\eta_3)\cr {\cal B}&=&
a_{12}=a_{14}=a_{32}=a_{34}=-{1\over 2}{{g_A}\over {g_B}} \gamma_1
(1-\gamma_2 \eta_2)\cr {\cal C}&=&
a_{13}=a_{31}=a_{24}=a_{42}=-\gamma_2(1-\gamma_1 \eta_1) \cr {\cal
D} &=& a_{21}=a_{23}=a_{41}=a_{43}=-{1\over 2}{{g_B}\over {g_A}}
\gamma_1 (1-\gamma_2 \eta_2)
\end{eqnarray}
and
\begin{equation}
{\cal N}(T)={N\over 4} {{\mu _0^2 H_0}\over {k T}} \Big[
1-\gamma_1 \eta_1 -\gamma_2 \eta_2 +{1\over 2} \gamma_2 \gamma_1
(\eta_1+\eta_3) \Big]
\end{equation}

\begin{figure}
\includegraphics[width=12cm,angle=0,bbllx=147pt,bblly=213pt,%
bburx=488pt,bbury=693pt,clip=true]{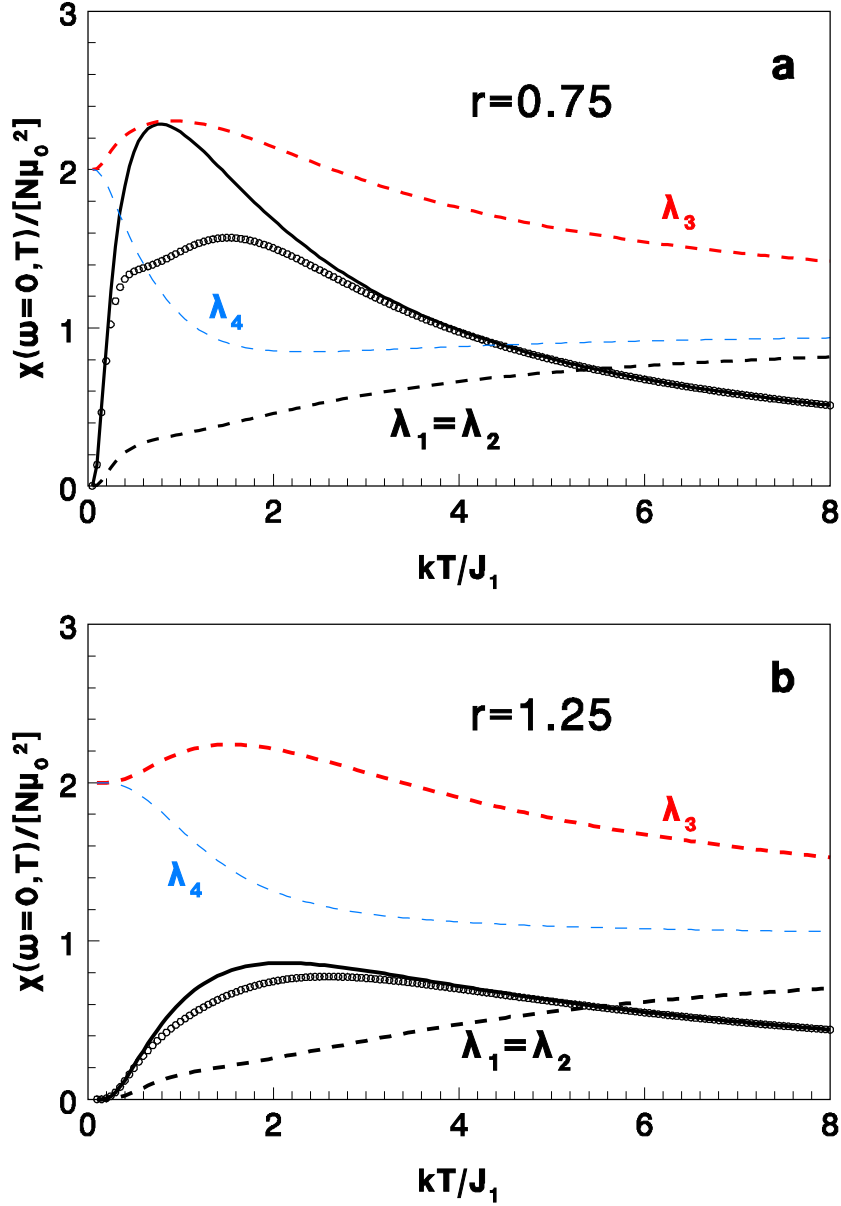} \caption{(color
online) Temperature dependence of the static susceptibility
$\chi(\omega=0,T)$ of an ANNNI chain with $J_1=1$, $g_A=g_B=2$,
for two different values of the nnn exchange constant: (a)
$J_2=-0.75$ and (b) $J_2=-1.25$, corresponding to competition
ratio ${1\over 2}<r<1$ and $r>1$ respectively (strong nnn
antiferromagnetism). The thick line is the exact transfer matrix
result, while open circles denote the approximate calculation,
Eq.~(\ref{susc4}). The temperature dependence of the eigenvalues
$\lambda_1=\lambda_2$, $\lambda_3$ and $\lambda_4$ (\ref{eigen4})
is also shown by the dashed lines.}
\end{figure}

Diagonalizing the matrix of coefficients, the time dependence of
the eigenmodes is found to be ($n=1,2,3,4$)
\begin{equation}
\label{generalANNNI} {\cal M}_n(t)={\cal
M}_n(t_0)e^{-{{t-t_0}\over {\tau_i}}}+{\cal N}(T) f_n\int_{t_0}^t
dt^{\prime} e^{{{t^{\prime}-t}\over {\tau_i}}} e^{-i \omega
t^{\prime}}
\end{equation}
\noindent where $\tau_n={1\over {\alpha \lambda_n}}$ are the
relaxation times and $f_1=f_2=0$, $f_3={{g_B}\over 2}(g_B-g_A)$,
$f_4={{g_B}\over 2} (g_B+g_A)$. The eigenvalues ($\lambda_n$) of
the $4 \times 4$ nonsymmetric matrix of the coefficients turn out to be
independent of $g_A$ and $g_B$
\begin{eqnarray}
\label{eigen4} \lambda_1&=& \lambda_2=1+\gamma_2(1-\gamma_1
\eta_1)+{1\over 2} \gamma_2 \gamma_1 (\eta_1 +\eta_3)\cr
\lambda_3&=& 1+\gamma_1 (1-\gamma_2 \eta_2)-\gamma_2(1-\gamma_1
\eta_1)+{1\over 2} \gamma_2 \gamma_1 (\eta_1 +\eta_3) \cr
\lambda_4&=&1-\gamma_1 (1-\gamma_2 \eta_2)-\gamma_2(1-\gamma_1
\eta_1)+{1\over 2} \gamma_2 \gamma_1 (\eta_1 +\eta_3)
\end{eqnarray}

For non vanishing magnetic field, the time dependence of the
eigenmodes is
\begin{equation}
{\cal M}_n(t)={\cal N}(T) {{f_n}\over {\lambda_n}}{{1}\over {1-i
\omega \tau_n}} e^{- i \omega t}~~~~(n=1,2,3,4)
\end{equation}

The relationships between the eigenmodes ${\cal M}_n(t)$ and the
sublattice magnetizations $M_n(t)$ ($n=1,2,3,4$) are
\begin{eqnarray}
{\cal M}_1(t)&=&{1\over 2} \Big[ M_4(t)-M_2(t) \Big] \cr {\cal
M}_2(t)&=&{1\over 2} \Big[ M_3(t)-M_1(t) \Big] \cr {\cal
M}_3(t)&=&{1\over 4} \Big[ \Big( M_4(t)+M_2(t) \Big) -{{g_B}\over
{g_A}} \Big(M_3(t)+M_1(t)\Big) \Big]\cr {\cal M}_4(t)&=&{1\over 4}
\Big[ \Big( M_4(t)+M_2(t) \Big) +{{g_B}\over {g_A}}
\Big(M_3(t)+M_1(t)\Big) \Big]
\end{eqnarray}
and conversely
\begin{eqnarray}
M_1(t)&=&{{g_A}\over {g_B}} \Big[ {\cal M}_4(t)-{\cal M}_3(t)
\Big]-{\cal M}_2(t)\cr M_2(t)&=&{\cal M}_4(t)+{\cal M}_3(t)-{\cal
M}_1(t)\cr M_3(t)&=&{{g_A}\over {g_B}}\Big[ {\cal M}_4(t)-{\cal
M}_3(t) \Big]+{\cal M}_2(t)\cr M_4(t)&=&{\cal M}_4(t)+{\cal
M}_3(t)+{\cal M}_1(t)
\end{eqnarray}

The total magnetization is
\begin{equation}
M_{tot}(t)=\sum_{i=1}^4 M_i(t) =2{{g_B+g_A}\over {g_B}} {\cal
M}_4(t)+2{{g_B-g_A}\over {g_B}} {\cal M}_3(t)=\chi(\omega)
H_0 e^{-i \omega t},\end{equation} \noindent where the complex
susceptibility $\chi(\omega,T)$ is given by
\begin{eqnarray}
\label{susc4} \chi(\omega,T)&=&N ~{{\mu_0^2}\over {k T}}~\Big[
1-\gamma_1 \eta_1 -\gamma_2 \eta_2 +{1\over 2} \gamma_2 \gamma_1
(\eta_1+\eta_3) \Big]\cr &\times& \Bigg[\Big({{g_B+g_A}\over
2}\Big)^2 {1\over {\lambda_4}} {1\over {1-i\omega \tau_4}}+
\Big({{g_B-g_A}\over 2}\Big)^2 {1\over {\lambda_3}} {1\over
{1-i\omega \tau_3}}\Bigg].
\end{eqnarray}

\noindent The approximate static susceptibility, calculated from
Eq.~(\ref{susc4}) for zero frequency, is shown in Fig.~4a for
${1\over 2}<r<1$. One immediately notices that, in striking
contrast with the case $0<r<{1\over 2}$ displayed in Fig.~3, the
low temperature behavior of the static susceptibility is correctly
reproduced.

The latter feature appears at odds with the expectation that a
decoupling approximation should work better the higher the
temperature. However it is worth noticing that, for the 1D ANNNI
model, the $T \to 0$ asymptotic behavior of the {\it static}
two-spin correlation functions is very different depending on the
value of $r$. For $0<r<{1\over 2}$ both the inter-  and the
intra-sublattice spin-spin correlations are strong ($\eta_1
\approx \eta_2 \approx \eta_3 \approx 1$, see Note
\onlinecite{notaeta} later). In contrast, for $r>{1\over 2}$ the
intersublattice correlations are strong ($\eta_2 \approx - 1$, see
Eq.~(\ref{expansions}) later), whereas the intrasublattice
correlations are exponentially vanishing ($\eta_1 \approx \eta_3
\approx 0$, see Eq.~(\ref{expansions})). At intermediate
temperatures intrasublattice correlations become significant, too,
and the decoupling approximation becomes less satisfactory; at
high temperatures, it works well again, since all correlations
(both intra- and inter-sublattice) decrease.

It should be remarked that the above considerations about the
behavior of {\it static} correlation functions can not, on their
own, account for the good agreement found, at low $T$, in the case
${1\over 2}<r<1$. In fact, the use of equilibrium values for the
spin correlations might be questionable, since the $T=0$ Glauber
dynamics does not lead to the ground state of the 1D ANNNI model
in the entire region $0<r<1$.\cite{Redner} To this regard, first
we observe that the physical mechanism which at $T=0$ prevents the
system from reaching the ground state is different, for ${1\over
2}<r<1$, with respect to the case $0<r<{1\over
2}$.\cite{Redner,Sen,Sen2} Next, considering that at $T=0$ a 1D
model is simultaneously in the ordered phase {\it and} at its
critical point, while our theory applies at $T>0$, we believe that
some insight into the problem might be provided by a careful study
of the role of a small but non-zero temperature on the coarsening
of the 1D ANNNI model.\cite{Bray,notaBray}

In Fig.~4b the approximate static susceptibility, calculated from
Eq.~(\ref{susc4}) for zero frequency in the case $r>1$, is
reported. A nice overall agreement with the exact transfer matrix
result\cite{Stephenson,Harada} is obtained. In this case our
approximate results are expected to be quite reliable since the
long-time approximation is well founded (for $r>1$, the static
equilibrium state can asymptotically be reached even at
$T=0$,\cite{Redner} and thus the use of {\it static} spin-spin
correlation functions is justified); moreover, the decoupling
approximation is expected to be satisfactory both at high and low
temperatures. Finally it is worth mentioning that, in the limiting
case $1/r=0$ ({\it i.e.}, $J_1=0$), the transfer matrix result for
the static susceptibility is exactly reproduced by
Eq.~(\ref{susc4}) for $\omega=0$ (not shown).

In Fig. 5 the temperature dependence of the amplitude of the
complex susceptibility $\vert \chi(\omega,T) \vert$, obtained from
Eq.~(\ref{susc4}), of an ANNNI chain with nnn antiferromagnetic
coupling dominating over the nn ferromagnetic one (competition
ratio $r=1.25$) is reported - for selected values of the
oscillation frequency $\omega$ of the external magnetic field -
both in the compensated ($g_A=g_B=2$) and uncompensated ($g_A \ne
g_B$) case. No resonant behavior was observed even in the
uncompensated case since, in the $T \to 0$ limit, both the
zero-field static susceptibility and the relaxation times
($\tau_3$ and $\tau_4$ in Eq. (\ref{susc4})) fail to diverge.
Thus, for low frequency, a resonance condition - similar to the
one in Eq.~(\ref{resonance_condition}) - can not be fulfilled. In
the case ${1\over 2}<r<1$ a qualitatively similar behavior for
$\vert \chi(\omega,T) \vert$ was found (not shown).

\begin{figure}
\includegraphics[width=16cm,angle=0,bbllx=25pt,bblly=290pt,%
bburx=520pt,bbury=596pt,clip=true]{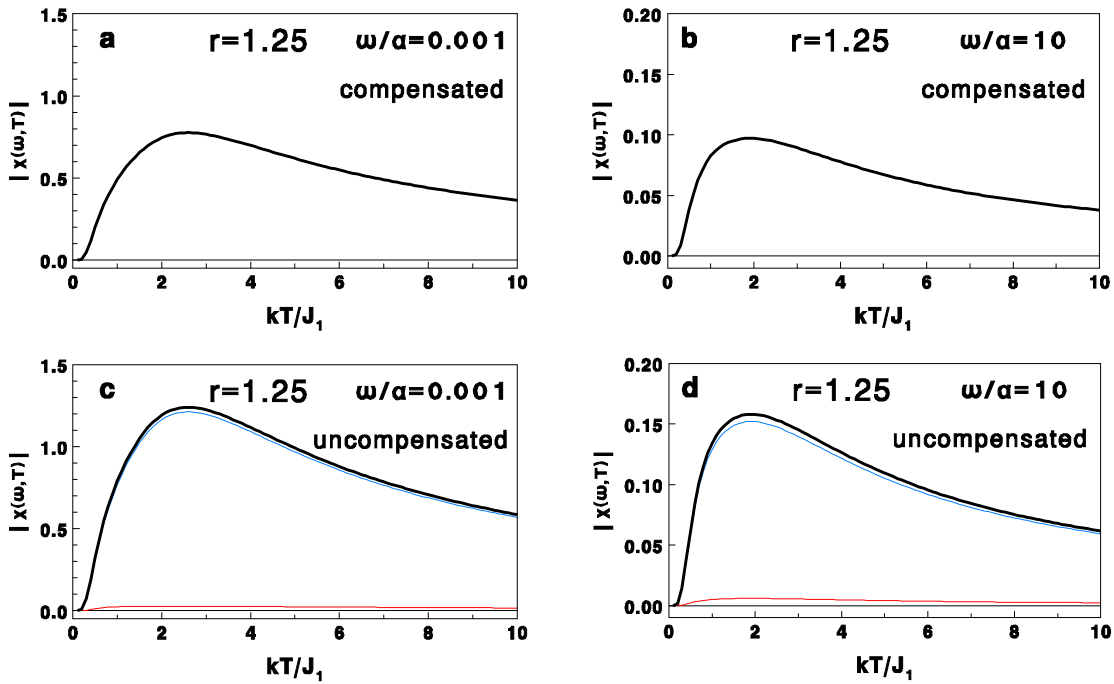} \caption{(color
online) Temperature dependence of the amplitude of the complex
susceptibility $\vert \chi(\omega,T) \vert$ for an ANNNI chain
with $J_1=1$, $J_2=-1.25$ ($r=1.25$), subject to a weak external
magnetic field oscillating at frequency $\omega$. Figures (a), (b)
refer to the compensated case ($g_A=g_B=2$), while Figures (c) and
(d) to the uncompensated case ($g_A=2,~g_B=3$), for selected
values of the frequency ($\omega/ \alpha=0.001$ and $10$). In
Figs.~(c,d) the thin (color) lines represent the contributions to
the amplitude of the two terms on the r.h.s. of Eq.~(\ref{susc4}),
while the thick (black) line is their sum. No resonant behavior is
observed.}
\end{figure}

\subsection{Critical dynamics of the 1D ANNNI model for $r>1$}

The identification of $r=0$, $r=1$ and $1/r=0$ as dynamic critical
transition points for the 1D ANNNI model with single spin-flip
Glauber dynamics was recently proposed in theoretical studies of
$T=0$ coarsening\cite{Redner} ({\it i.e.}, the relaxation of the
system into the ground state after a quench from high temperature)
and $T=0$ persistence\cite{Sen} ({\it i.e.}, the probability for a
spin to remain in its original state after a quench from high
temperature).  In such $T=0$ studies, the dynamic critical
exponent $z$ is customarily defined as the inverse of the growth
exponent $n$ of the domain size
\begin{equation}
\label{zetaprimo} L(t) \simeq t^n \simeq t^{1/z^{\prime}}.
\end{equation}
For the 1D nn Ising model, analytical calculations\cite{Bray}
provided $z^{\prime}=2$. For the 1D ANNNI model with $r>1$
numerical calculations\cite{Sen,notaSen} predicted a somewhat
higher dynamic exponent, $z^{\prime} \simeq 2.3$. Finally, it is
worth noting that Sen and Dasgupta,\cite{Sen} in their study of
$t=0$ persistence in the ANNNI chain, found that the dynamic
critical exponent $z^{\prime}$ undergoes abrupt changes for $r=0$
(when a slight amount of nnn interaction is added to the nn one),
for $1/r=0$ (when a slight amount of nn interaction is added to
the nnn one), as well as for $r=1$.\cite{Sen}

The fair accuracy of our approximate theoretical approach in
describing the low temperature static susceptibility of the ANNNI
chain with $r>1$, see Fig. 4b, encouraged us to tentatively
estimate the dynamic critical exponent. However, since we work at
finite temperature, rather than at $T=0$, we use a different
definition, namely\cite{daSilva,Luscombe}
\begin{equation}
\label{cexp} {1\over {\alpha \tau_1}}=\lambda_1 \approx
\Big({{1}\over {\xi}}\Big)^z
\end{equation}
where $\lambda_1$ is the smallest eigenvalue of the dynamical
matrix, see Eq.~(\ref{eigen4}), and $\xi$ is the static
correlation length of the infinite system (the lattice constant
$c$ along the chain was set to 1). For the compensated case
$g_A=g_B$, the latter quantity can be analytically calculated
using the transfer matrix method,\cite{Stephenson} see
Eq.~(\ref{xiHarada}), and for $r>1$ its expansion in the $T \to 0$
limit turns out to be
\begin{equation}\Big( {{1}\over {\xi}}\Big)^2 \approx
{1\over 4}e^{{{2(J_1-2\vert J_2 \vert) }\over {kT}}},
\end{equation}
where we have explicitly taken into account that $J_1>0$ and
$J_2<0$.

Taking into account the $T\to 0$ asymptotic behavior, for $r>1/2$, \cite{notaeta}
of the $\gamma_i$ and $\eta_i$ coefficients \begin{eqnarray}\label{expansions} && \gamma_1
\approx 1-2e^{-{{4J_1}\over {kT}}}~~~\gamma_2\approx
-1+2e^{-{{4\vert J_2\vert}\over {kT}}} \cr &&\eta_1\approx {1\over
2} e^{{{J_1-2\vert J_2 \vert}\over {k T}}}~~~\eta_2\approx
-1+e^{{{J_1-2\vert J_2 \vert}\over {k T}}} ~~~\eta_3\approx
-{3\over 2} e^{ {{J_1-2\vert J_2 \vert}\over {k T}} },
\end{eqnarray}
\noindent for the inverse of the longest relaxation time we
obtain, provided that $J_1 \ne 0$
\begin{equation}
\label{relaxtime} {1\over {\alpha\tau_1}}=\lambda_1 \approx
e^{{{J_1-2\vert J_2 \vert }\over {kT}}}.
\end{equation}
\noindent In the special case $J_1=0$ ({\it i.e.}, $1/r=0$),
letting $\gamma_1=0$ in Eq.~(\ref{eigen4}) and using the $T \to 0$
expansion for $\gamma_2$ in Eq.~(\ref{expansions}), we obtain
\begin{equation}
{1\over {\alpha\tau_1}}=\lambda_1 \approx 2 e^{-{{4\vert J_2 \vert
}\over {kT}}}.
\end{equation}

In conclusion, within our approximate theoretical scheme, the
dynamic critical exponent of the 1D ANNNI chain with competing nn
and nnn exchange interactions was found to be $z=1$ for any finite
$r>1$, while in the absence of competing interactions ({\it i.e.},
for $r=0$ and  $1/r=0$) we found $z=2$. Notice that, for the 1D
Ising model with exchange limited to the nn ($r=0$), the value
$z=2$, obtained using the definition in
Eq.~(\ref{cexp}),\cite{Luscombe} coincides with the value
$z^{\prime}=2$, obtained using the definition in
Eq.~(\ref{zetaprimo}).\cite{Bray} This appears {\it not} to be the
case for the 1D ANNNI model with $1<r<+\infty$, where the values
$z=1$ (present work) and $z^{\prime}\simeq 2.3$ (References
\onlinecite{Sen,notaSen}) were found. In order to ascertain the
origin of this discrepancy, we believe that it would be useful to
study the role of a small but non-zero temperature ($T>0$) on the
coarsening dynamics of the 1D ANNNI model.\cite{notaBray}

\section{Conclusions}

In conclusion, in this paper we have studied the effect of
antiferromagnetic interactions on the single spin-flip Glauber
dynamics of two different one-dimensional (1D) Ising models with
spin $\pm 1$. For the first model, an Ising chain with
antiferromagnetic exchange interaction limited to nearest
neighbors and subject to an oscillating magnetic field, the system
of master equations describing the time evolution of sublattice
magnetizations can easily be solved within a linear field
approximation and a long time limit. Resonant behavior of the
magnetization as a function of temperature (stochastic resonance)
is found, at low frequency, only when spins on opposite
sublattices are uncompensated owing to different gyromagnetic
factors ({\it i.e.}, in the presence of a ferrimagnetic short
range order). For the second model, the axial next-nearest
neighbor Ising (ANNNI) chain, where the  nnn antiferromagnetic
exchange coupling is assumed to compete with the nn ferromagnetic
one, the long time response of the model to a weak, oscillating
magnetic field is investigated in the framework of a decoupling
approximation for three-spin correlation functions, which is
required to close the system of master equations. Within such
approximate theoretical scheme, the $T=0$ dynamics of the
Ising-Glauber chain with competing interactions is found to be in
a different universality class than that of the Ising chain with
antiferromagnetic exchange limited to nearest neighbors ($r=0$) or
limited to next-nearest neighbors ($1/r=0$). In particular, we
find an abrupt change in the $T=0$ dynamic behavior of the model
in the neighborhood of the dynamic critical point $1/r=0$ since,
when a slight amount of ferromagnetic nn exchange is added to the
antiferromagnetic nnn exchange, we find that the critical exponent
$z$, defined by Eq.~(\ref{cexp}), changes abruptly from $z=2$ to
$z=1$. Considering that $z=2$ is also the value of the dynamic
critical exponent for the unfrustrated nn Ising chain, one might
expect similar abrupt changes in $z$ to occur also in the
neighborhood of the dynamic critical points $r=0$ ({\it i.e.} when
a slight amount of AF nnn exchange is added to the nn F exchange)
and $r=1$, as suggested by studies of $T=0$ coarsening
dynamics\cite{Redner} and $T=0$ persistence\cite{Sen} in the ANNNI
chain. Unfortunately, the inaccuracy of our approximate
theoretical scheme in reproducing the static susceptibility of the
1D ANNNI model with $0<r \le 1$ for low temperature prevented us
from calculating the dynamic critical exponent in this range of
the competition ratio.

\appendix
\section{Analytic transfer matrix results for the static
properties of 1D Ising models}

\subsection{The 1D nearest neighbor Ising model with alternating spins in a static field}

In this subsection we calculate, within the transfer matrix
formalism,\cite{Stanley} the static properties of the 1D Ising
model, Eq.~(\ref{ferrimagnet}), with nearest neighbor coupling $J$
of either sign, subject to a static magnetic field $H$ ({\it
i.e.}, $\omega=0$). Two types of spins with different gyromagnetic
factors ($g_A \ne g_B$) are assumed to alternate along the chain.
Taking periodic boundary conditions, the partition function of the
chain of length $N$ (with $N$ even without loss of generality) can
be expressed as
\begin{equation}
Z_N={\rm Tr}\big( e^{-{{\cal H}\over {kT}}}\big)=\sum_{\sigma_1=\pm 1}
\sum_{\sigma_2=\pm 1}\cdots \sum_{\sigma_N=\pm 1}
K(\sigma_1,\sigma_2) L(\sigma_2,\sigma_3) \cdots
K(\sigma_{N-1},\sigma_N) L(\sigma_N,\sigma_1)
\end{equation}
\noindent where, letting ${\cal J}=J/(k T)$, $h_A=(g_A \mu_0 H_0)/(k T)$,
$h_B=(g_B \mu_0 H_0)/(k T)$, the two different kernels $K$ and $L$ are defined as
\begin{equation}
K(\sigma_{2i-1},\sigma_{2i})=e^{{\cal J}\sigma_{2i-1}
\sigma_{2i}}e^{{1\over 2}(h_A \sigma_{2i-1} +h_B\sigma_{2i})}~~~~
L(\sigma_{2i},\sigma_{2i+1})=e^{{\cal J} \sigma_{2i}
\sigma_{2i+1}}e^{{1\over 2}(h_B \sigma_{2i} +h_A \sigma_{2i+1})}.
\end{equation}
Summing over the even sites, $Z_N$ can be expressed as
\begin{equation}
Z_N=\big(\Lambda_+\big)^{N\over 2}+\big(\Lambda_-\big)^{N\over 2}
\end{equation}
\noindent in terms of the eigenvalues
\begin{eqnarray}
\Lambda_{\pm}&=&e^{2{\cal J}}\cosh(h_A+h_B)+e^{-2{\cal
J}}\cosh(h_A-h_B)\pm \cr\cr &\pm& \sqrt{e^{4{\cal
J}}\cosh^2(h_A+h_B)+e^{-4{\cal J}}\cosh^2(h_A-h_B)
+2\cosh(h_A+h_B)\cosh(h_A-h_B)+2-e^{4{\cal J}}-e^{-4{\cal J}} }
\end{eqnarray}
\noindent of the real symmetric $2 \times 2$ matrix
\begin{equation} S=\left(
\begin{array}{cc} e^{2{\cal J}+h_A+h_B}+e^{-2{\cal J}+h_A-h_B}
 &  e^{h_B}+e^{-h_B} \\
e^{h_B}+e^{-h_B} &  e^{2{\cal J}-h_A-h_B}+e^{-2{\cal J}-h_A+h_B}
 \end{array} \right).
\end{equation}
It is immediate to verify that, in the limit $g_A=g_B$, the
well-known result for the 1D nn Ising chain in a static external
field is recovered.\cite{Stanley} In the thermodynamic limit $N
\to \infty$, only the larger eigenvalue $\Lambda_+$ matters, $Z_N
\to (\Lambda_+)^{N\over 2}$, and the static susceptibility in zero
field can be expressed in terms of its second derivative with
respect to the field $H$
\begin{equation}
\label{chisAF} \chi(\omega=0,T)={N\over 2} kT \Big[ {1\over
{\Lambda_+}} {{\partial^2 \Lambda_+}\over {\partial H^2}}
\Big]_{H=0} = {{N \mu_0 ^2}\over {k T}} \Big[ \Big({{g_A+g_B}\over
2}\Big)^2 e^{{2J}\over {kT}}+ \Big({{g_A-g_B}\over 2}\Big)^2
e^{-{2J}\over {kT}} \Big].
\end{equation}

\subsection{The 1D ANNNI model in zero field}

In this subsection we collect, for the reader's convenience, some
exact results for the static properties of the 1D ANNNI model in
zero field, Eq.~(\ref{ANNNI}), which were obtained by
Stephenson\cite{Stephenson} and Harada\cite{Harada} in the case of
a linear chain with $N$ identical spins ($g_A=g_B=g$ and
$\sigma=\pm 1$). Using the transfer matrix method, the partition
function can be exactly expressed as
\begin{equation}
\label{TM} Z_N=(\lambda_+)^N+(\lambda_-)^N
\end{equation}
in terms of the eigenvalues of the symmetric $2 \times 2$ matrix
\begin{equation}
S=\left(\begin{array}{cc} a & c
\\ c& b \end{array}\right)=\left(
\begin{array}{cc} e^{{{J_2+J_1}\over {k T}}}
 & e^{{{J_2-J_1}\over {k T}}} \\
e^{{{J_2-J_1}\over {k T}}} & e^{-{{J_2}\over {k T}}}  \end{array}
\right)
\end{equation}
The eigenvalues take the form\cite{Stephenson}
\begin{equation}
\lambda_{\pm}={1\over 2}[a+b\pm \Delta]=e^{{{J_2}\over {k
T}}}\Big[ \cosh\Big({{J_1}\over {k T}} \Big)\pm
\sqrt{\sinh^2\Big({{J_1}\over {k T}} \Big)+e^{-{{4J_2}\over {k
T}}}}\Big]
\end{equation}
where
\begin{equation}\Delta=\sqrt{(a-b)^2+4c^2}=2 e^{ {{J_2}\over
{k T}} } \sqrt{ \sinh^2\Big({{J_1}\over {k
T}}\Big)+e^{-{{4J_2}\over {k T}}} }.\end{equation} Both
$\lambda_{+}$ and $\Delta$ are always real positive quantities.

In the thermodynamic limit $N \to \infty$, the static two spin
correlation function $\eta_n$ take the form\cite{Stephenson}
\begin{equation}
\eta_n=\langle\sigma_j\sigma_{j+n}\rangle={1\over {2
\big(\lambda_+\big)^n}} \Big[\big(\mu_+\big)^n \Big(\
1+{{a^2-b^2}\over {\Delta \Delta^{\prime}}}\Big)
+\big(\mu_-\big)^n \Big(\ 1-{{a^2-b^2}\over {\Delta
\Delta^{\prime}}}\Big)\Big]
\end{equation}
where the quantities $\Delta^{\prime}$, defined as
\begin{equation}
\Delta^{\prime}=\sqrt{(a+b)^2-4c^2}=2 e^{ {{J_2}\over {k T}} }
\sqrt{ \cosh^2\Big({{J_1}\over {k T}}\Big)-e^{-{{4J_2}\over {k
T}}} },
\end{equation} and
\begin{equation}\mu_{\pm}=e^{{{J_2}\over {k
T}}}\Big[ \sinh\Big({{J_1}\over {k T}} \Big)\pm
\sqrt{\cosh^2\Big({{J_1}\over {k T}} \Big)-e^{-{{4J_2}\over {k
T}}}}\Big]\end{equation} may be complex. More precisely, the
quantities $\mu_{\pm}$ are real for $T<T_D$ and complex conjugates
for $T>T_D$. $T_D$ is the so-called disorder point, defined by the
equation $\Delta^{\prime}(T_D)=0$, which has solutions for
$0<r<1/2$ at some finite temperature $T_D$. For $T<T_D$ the static
equilibrium two-spin correlation functions $\eta_n=\langle
\sigma_j \sigma_{j+n} \rangle$ present a monotonic exponential
decay, while for $T>T_D$ they have an oscillating exponential
decay.\cite{Stephenson}

Summing over all pair correlations, the exact zero field static
susceptibility can be expressed as\cite{Stephenson}
\begin{equation}
\label{suscSteph}
\chi(\omega=0,T)=\Big({{{g^2 \mu_0}^2}\over {k T}} \Big) \Big(
{{a+b}\over {\Delta}}\Big) \Big[ {{a(a-b+\Delta)+2c^2}\over
{b(b-a+\Delta)+2c^2}}\Big].
\end{equation}

The wave-vector dependent susceptibility, defined as
\begin{equation}
\chi(q)=N {{g^2 \mu_0^2 }\over {k T}} \sum_n \langle \sigma_j
\sigma_{j+n} \rangle e^{i q n}
\end{equation}
presents a maximum at a wave-vector $q_m$, which is given by\cite{Harada}
\begin{equation}
\cos q_m={{(\mu_+ +\mu_-)(\lambda_+ -\lambda_-)}\over {4\mu_+
\mu_-}}
\end{equation}
For $0<r<1/4$ one has $q_m=0$ at all temperatures ,
while for $1/4<r<1/2$ there is a definite temperature $T_L$ ($\ne
T_D$) above which $q_m \ne 0$, whereas for $T<T_L$ $q_m = 0$. When
$1/2<r$, one has $q_m(T=0)=\pi/2$. In the limit of $T\to \infty$,
$q_m$ tends to the mean field value $\cos q_m=1/(4r)$.
Expanding $\chi(q)$ in the neighborhood of $q_m$ up to second
order in $\Delta q=q_m-q$, one obtains a Lorentzian form, and the
correlation length $\xi$ can be defined in terms of its full width
at half maximum as
\begin{equation}
\chi(q)={{\chi(q_m)}\over {1+\xi^2 (\Delta q)^2}}
\end{equation}
\noindent and turns out to be
\begin{eqnarray}
\label{xiHarada} \Big({1\over {\xi}}\Big)^2&=& {{(\lambda_+ -
\lambda_- - \mu_+ -\mu_- )^2}\over {(\mu_+ + \mu_-)(\lambda_+ -
\lambda_-)-4\mu_+ \mu_-}} ~~~{\rm for}~q_m=0, \cr \Big({1\over
{\xi}}\Big)^2&=& {{(\mu_+ -\mu_-)^2(\lambda_+ + \lambda_-)^2}\over
{(\mu_+ + \mu_- )^2(\lambda_+ - \lambda_-)^2-16\mu_+^2\mu_-^2}}
~~~{\rm for}~q_m \ne 0.
\end{eqnarray}
For $1/4<r<1/2$, it turns out that at $T_L$ the correlation length
becomes zero, which is a characteristic of the Lifshitz
point.\cite{RednerStanley}


\begin{thebibliography}{99}
\bibitem{notaBray}
While for the 1D nn Ising model such a study was performed by
Bray,\cite{Bray} for the 1D ANNNI model it has not yet been
performed, to our knowledge, and is deferred to future work.
\bibitem{notaSen}
It is worth noting that such a value, $n={1\over {2.3}}$, for the
growth exponent turns out to be in close agreement with previous
analytical predictions, obtained by Redner and
Krapivsky\cite{Redner} in the same regime of strong nnn
antiferromagnetic exchange. In fact for $r>1$ the latter authors
estimated the density of domains of length 3 and 1 to scale as
$A(t) \simeq \sqrt{{{\ln t}\over {t}}}$ and $B(t) \simeq {1\over
\sqrt{{t \ln t}}}$, respectively. For $t \gg 1$, it turns out that
the average of $A(t)$ and $B(t)$ (which is inversely proportional
to the domain size $L(t)$) scales just as ${1\over {t^{1/2.3}} }$.
\bibitem{notaeta}
In the case of ferromagnetic ground state, $0<r<1/2$, the $T\to 0$
asymptotic behavior of the correlation functions for the 1D ANNNI
model in zero field is instead: $\eta_1=1-2e^{{{-2J_1+4\vert J_2
\vert}\over {kT}}}$, $\eta_2=1-4e^{{{-2J_1+4\vert J_2 \vert}\over
{kT}}}$, $\eta_3=1-6e^{{{-2J_1+4\vert J_2 \vert}\over {kT}}}$.
\end{thebibliography}
\end{document}